\documentclass[english,aps,twocolumn,showpacs,superscriptaddress]{revtex4-1}
\usepackage[T1]{fontenc}
\usepackage[latin1]{inputenc}
\usepackage{dvipost}
\usepackage{color}
\usepackage{graphicx}
\usepackage{epsfig}
\usepackage{epstopdf}
\usepackage{amssymb}
\usepackage{color}
\makeatletter

\usepackage{dcolumn}

\usepackage{bm}
\usepackage{cancel}
\usepackage[normalem]{ulem}
\usepackage{amsfonts}
\usepackage{amsthm}
\usepackage{amssymb}
\usepackage{cancel}
\usepackage{graphicx}

\usepackage{babel}

\begin{document}

\title{\bf {The stabilizing effect of volatility in financial markets}}

\author{Davide Valenti}
\email{davide.valenti@unipa.it} \affiliation{Dipartimento di Fisica e Chimica,
Group of Interdisciplinary Theoretical Physics and CNISM, Universit\`{a} di
Palermo, Viale delle Scienze, edificio 18, I-90128 Palermo, Italy}

\author{Giorgio Fazio}
\email{giorgio.fazio@ncl.ac.uk}
\affiliation{Business School, Newcastle University, 5 Barrack Road, NE1 4SE, Newcastle upon Tyne, UK}
\affiliation{SEAS, Universit\`{a} di Palermo, Italy}

\author{Bernardo Spagnolo}
\email{bernardo.spagnolo@unipa.it} \affiliation{Dipartimento di Fisica e Chimica, Group
of Interdisciplinary Theoretical Physics and CNISM, Universit\`{a} di Palermo,
Viale delle Scienze, edificio 18, I-90128 Palermo, Italy}
\affiliation{Istituto Nazionale di Fisica Nucleare, Sezione di Catania, Italy}

\begin{abstract}
In financial markets, greater volatility is usually considered synonym of greater risk and instability. However, large market downturns and upturns are often preceded by long
periods where price returns exhibit only small fluctuations. To investigate this surprising feature, here we propose using the mean first hitting time, i.e. the average time a stock
return takes to undergo for the first time a large negative or positive variation, as an indicator of price stability, and relate this to a standard measure of volatility. In an empirical
analysis of daily returns for $1071$ stocks traded in the New York Stock Exchange, we find that this measure of stability displays nonmonotonic behavior, with a maximum, as a
function of volatility. Also, we show that the statistical properties of the empirical data can be reproduced by a nonlinear Heston model. This analysis implies that, contrary
to conventional wisdom, not only high, but also low volatility values can be associated with higher instability in financial markets.
\end{abstract}

\pacs{89.20.-a, 89.65.Gh,02.50.-r}

\keywords{financial instability, stock market volatility, noise enhanced stability, nonlinear Heston model}

\date{\today}
\maketitle

\begin{figure}[htbp]
\centering{\resizebox{5.7cm}{!}{\includegraphics{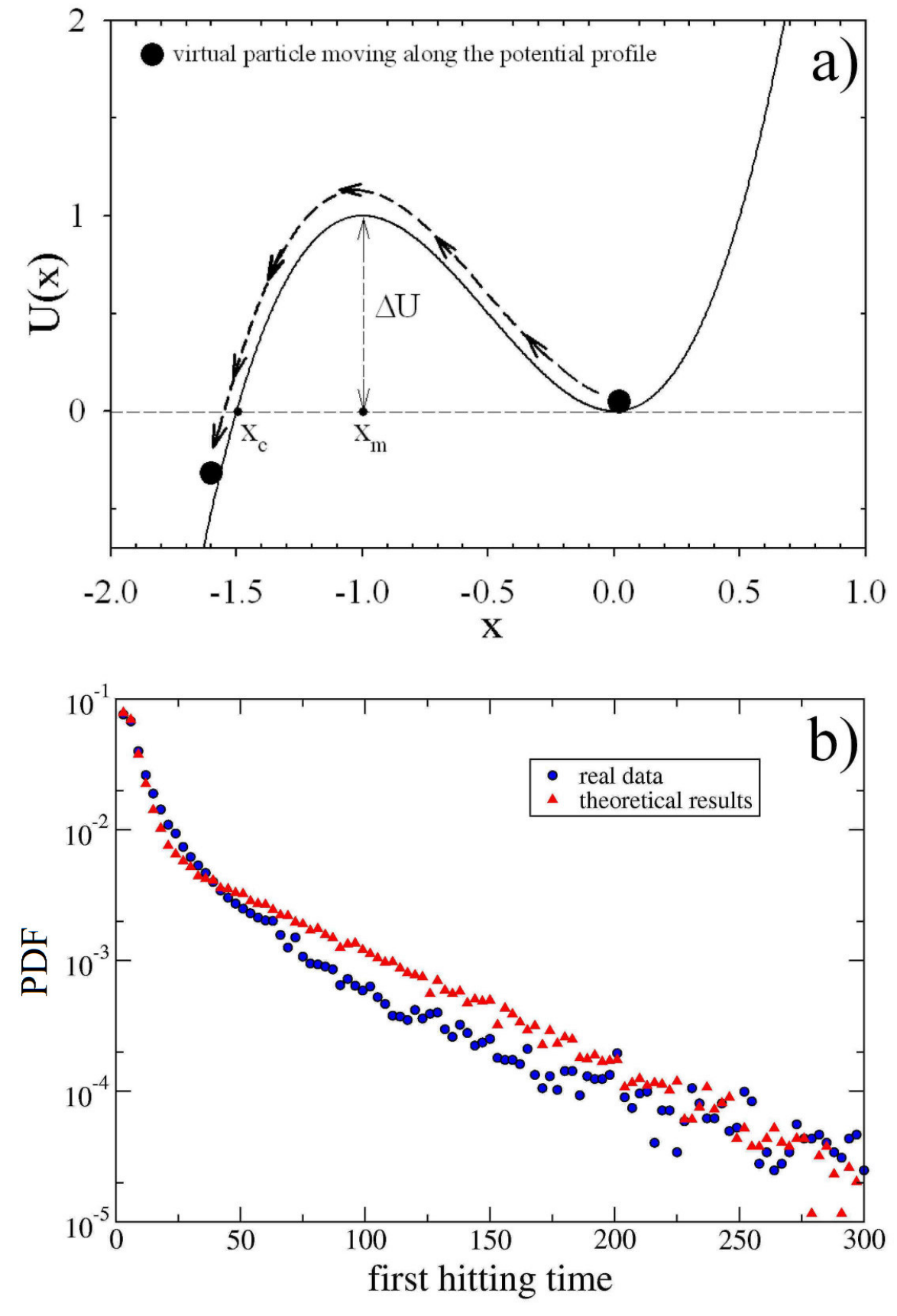}}}
\vspace{-0.3cm} \caption{a) Cubic potential used in the dynamic equation for the variable $x(t)$. The black circle denotes the starting position ($x_{0}=0.0$) used to obtain the
theoretical results. The potential parameters are $m = 2$ and $n = 3$. b) PDF of the first hitting times of the returns for real data (blue circles) and model (red triangles).}
\vspace{-0.12cm} \label{tau-ret_pdf-pot}
\end{figure}

\indent Volatility is typically considered a monotonic indicator of financial markets risk and instability. Recently, however, such a conventional wisdom has been questioned
by the observation that sizeable market downturns or upturns can be anticipated by periods of low volatility. Notable examples of this phenomenon include the 2008 financial
crisis, preceded by the so called "great moderation", and the Chinese crash in 2015. These episodes have received a lot of attention in the specialized press and have
popularized the so called Minsky's financial instability hypothesis~\cite{Min92} that periods of calm can project a false sense of security and lure agents into taking riskier
investment, preparing for a crisis~\cite{Yel14}. Therefore, a better characterization of the relationship between volatility and market stability seems particularly important.\\
\indent Searching for the empirical regularities and modeling complex market dynamics have typically been the objective of financial times series analysis, econophysics
and complex systems~\cite{Man95,Man00,Bou03,Ple03,Lil03,Bou08,Yak09}. In this literature, the importance of the statistical properties of volatility for portfolio optimization
strategies, risk management and financial stability have been underlined in Refs.~\cite{Wan08,Adr14,Eng04,Bia09}. Along these lines, investigations have looked at the
statistical properties of large volatilities~\cite{Yam05}, cross-correlations between volume change and price change~\cite{Pod09}, and temporal sequences of financial market
fluctuations around abrupt switching points~\cite{Pre11}. There, the authors argue that the end of microscopic or macroscopic trends in financial markets have a parallel with
metastable physical systems. Indeed, financial market stability is often associated with moderate levels of perceived uncertainty and measured-looking at the intensity of price
return fluctuations~\cite{Moh11,Dat10,Gad09} or stochastic volatility estimators based on first passage time statistics~\cite{And09}. However, both approaches cannot be
reconciled with the observed evidence discussed above.\\
\indent In this direction, a fundamental, and yet overlooked, question has to be addressed: what is the typical time scale before a large negative or positive stock return variation?
To answer this question, we propose exploiting the notions of ``level crossings'' and ``hitting times'' to monitor the stability of price returns and observing its relationship with
volatility~\cite{Bon07,Mas08,Note1}. In particular, the mean first hitting time (MFHT) or mean first passage time (MFPT), earlier introduced in~\cite{Note1,Note2,Kra40,Cha43,Fel51},
is the time it takes, on average, for a variable to cross for the first time a certain level, and it can provide the above mentioned time scale to observe modifications in market
scenarios. In finance~\cite{Note2}, the MFPT for mean-reverting processes was recently analyzed in ~\cite{Zha10,Mas12}.\\
\begin{figure*}[htbp]
\vspace{5mm}
\includegraphics[width=18cm,angle=0]{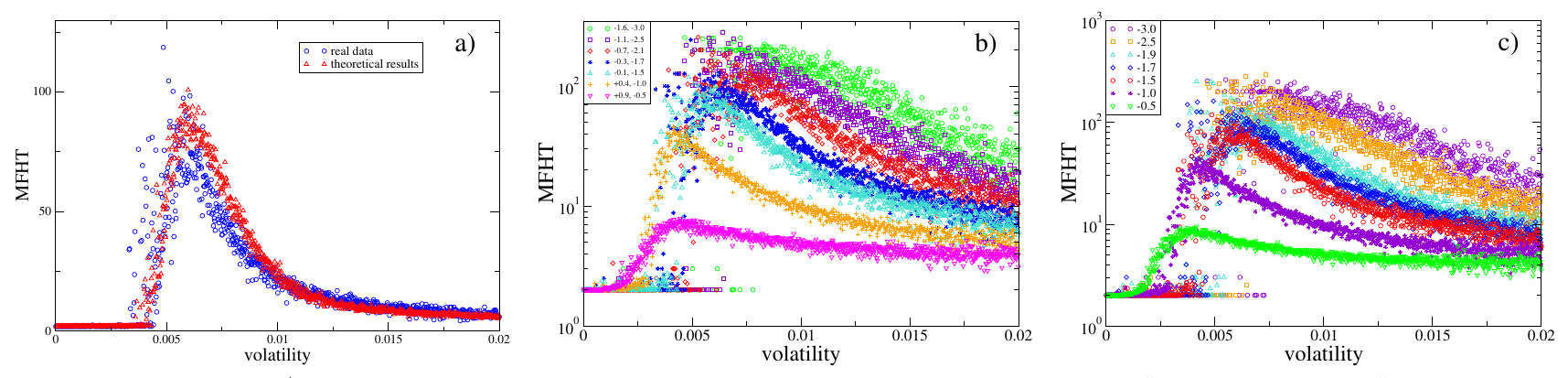}
\vspace{-0.4cm} \caption{Microcrash: a) MFHT as a function of volatility, with the thresholds $\Theta_i = - 0.1\bar{\sigma}^r$ and $\Theta_f = -1.5\bar{\sigma}^r$, from empirical time
series (blue circles) and theoretical results (red triangles), obtained from a nonlinear Heston model (Eqs.~(\ref{NLH})-(\ref{CIR})). b) MFHT vs volatility (real data) for fixed difference
between thresholds, $\Theta_f - \Theta_i = -1.4 \bar{\sigma}^r$, with $[\Theta_i, \Theta_f]$ ranging from $[+0.9\,\bar{\sigma}^r,\,-0.5\,\bar{\sigma}^r]$ to $[-1.6\,\bar{\sigma}^r,\,-3.0\,\bar{\sigma}^r]$;
c) MFHT vs volatility (real data) for fixed starting threshold $\Theta_i = - 0.1\bar{\sigma}^r$ and different final threshold $\Theta_f$, ranging from $-0.5\,\bar{\sigma}^r$ to $-3.0\,\bar{\sigma}^r$.}
\vspace{-0.3cm} \label{tau_crash}
\end{figure*}
\indent The MFHT is then utilized to measure the stability of price returns, defined as the resilience to large negative price variations: the longer this time, the more stable the series of
price returns~\cite{Bon07}. Observing the daily closing prices of a large number of stocks traded in the New York Stock Exchange (NYSE), we find that this measure of stability
has a nonmonotonic behavior, with a maximum, as a function of volatility. This result seems in line with the view discussed above that higher price return instability, corresponding
here to lower hitting times, is not only associated with high values but also with low values of volatility. As such, this measure can be considered as an important indicator of market
stability.\\
\indent Further, we are able to reproduce all the main statistical features of the price return dynamics of the considered stock market by using a nonlinear generalization of the
Heston model proposed in~\cite{Bon07}. \\
\indent We analyze the daily closing prices for 1071 stocks traded at the NYSE~\cite{Note3,Mic02,Bon03,Bon04,Note4}. In order to investigate episodes of instability associated with
large negative or positive return variations, following the literature on speculative pressure in the exchange market, we identify price changes as ``sizeable'' if they are larger than a
certain threshold, typically defined starting from the standard deviation~\cite{Eic95,Faz07}. In line with this literature, the robustness of the identification mechanism is also assessed
by considering different thresholds in the range of $1.5$ to $3$ times the standard deviation.\\
\indent In detail, we first transform the series of stock prices into daily returns, $r(t) = [(p(t) - p(t-1))/p(t-1)]$ and calculate the standard deviation $\sigma^r_i$ of each series over
the entire period. By averaging this quantity across all stocks in the sample, we obtain the overall volatility of the market over the observed period as
$\bar{\sigma}^r = \sum_{i=1}^N \sigma^r_i/N = 0.02254$, with $N = 1071$. We then proceed to compute the MFHT by considering the \emph{first hitting time} (FHT)~\cite{Note1},
and ensemble averaging over all FHTs measured in the price return series. This is the ``random time to hit'' for the first time the fixed final threshold,
$\Theta_f$, starting from a given initial position $\Theta_i$, where the two thresholds are defined, in line with~\cite{Eic95,Faz07}, starting from the market standard deviation,
as $\Theta_i=\theta_i \bar{\sigma}^r$ and $\Theta_f=\theta_f\bar{\sigma}^r$. The parameters $\theta_i$ and $\theta_f$ define the ``stability'' window and, therefore, how large a
variation has to be in order to determine an escape from a metastable state~\cite{Note4,Eic95,Faz07}. \\
\indent This allows us to obtain several subseries, each corresponding to one first hitting time. The standard deviation of each subseries gives the value of volatility $v$, corresponding to
each FHT. Averaging all the FHTs corresponding to the same volatility value yields the nonmonotonic behavior of the MFHT versus the volatility shown in Fig.~\ref{tau_crash}a (blue circles),
where the values of the threshold parameters are $\theta_i = -0.1$ and $\theta_f = -1.5$. In particular, we note that the MFHT takes smaller values for lower levels of volatility, i.e. the series
of returns exhibit negative jumps equal to or less than $-1.5\bar{\sigma}^r$, after short time intervals. This corresponds to a fast exit of the stock return from the fixed region $[\Theta_i,\, \Theta_f]$.
As volatility increases, the time spent within this region also increases, which is a signal that the market is becoming more stable. A further increase of volatility, however, shortens the MFHT,
and stability decreases. This implies that in the intermediate region we observe a \emph{stabilizing effect of volatility}. Fig.~\ref{tau_crash}a is a clear representation of the relation between MFHT,
i.e. the time returns stay within the fixed region, and the size of volatility $v$~\cite{Note5}. \\
\indent In order to cross-validate the robustness of this result, we have also investigated whether this effect persists for: (i) different thresholds with fixed interval size,
$\Theta_f - \Theta_i = -1.4 \bar{\sigma}^r$ (Fig.~\ref{tau_crash}b); (ii) fixed starting threshold $\Theta_i$ but different final thresholds $\Theta_f$ (Fig.~\ref{tau_crash}c). The results indicate
that the nonmonotonic behavior of the MFHT as a function of volatility is ``robust'' to sizeable variations of the two thresholds.\\
\begin{figure*}[htbp]
\vspace{5mm}
\includegraphics[width=18cm,angle=0]{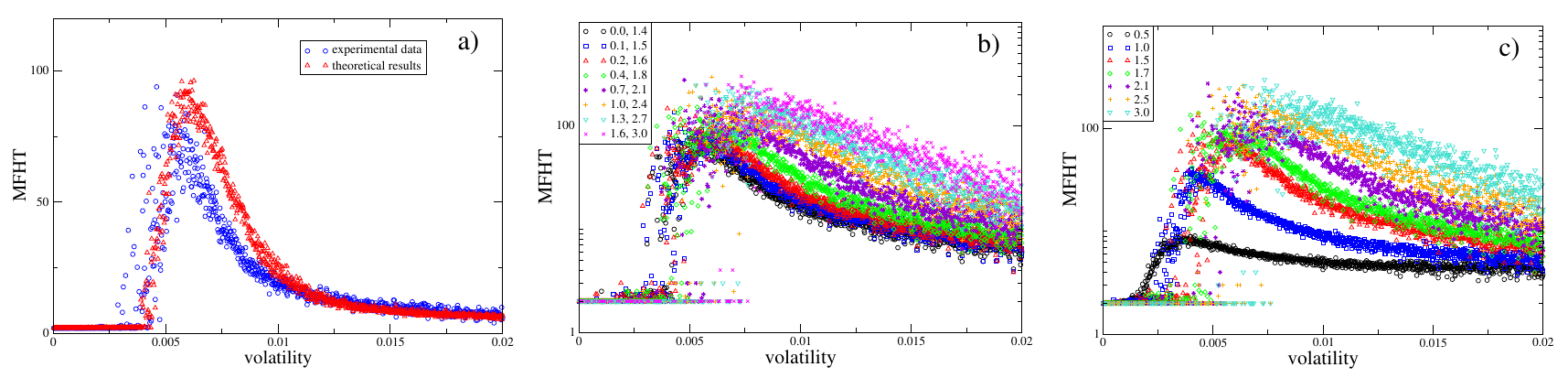}
\vspace{-0.4cm} \caption{Rally: a) MFHT as a function of volatility, with the thresholds $\Theta_i = + 0.1\bar{\sigma}^r$ and $\Theta_f = +1.5\bar{\sigma}^r$, from empirical time
series (blue circles) and theoretical results (red triangles), obtained from a nonlinear Heston model. b) MFHT vs volatility (real data) for fixed difference between
thresholds, $\Theta_f - \Theta_i = +1.4 \bar{\sigma}^r$, with $[\Theta_i, \Theta_f]$ ranging from $[-0.9\,\bar{\sigma}^r,\,+0.5\,\bar{\sigma}^r]$ to $[+1.6\,\bar{\sigma}^r,\,+3.0\,\bar{\sigma}^r]$;
c) MFHT vs volatility (real data) for fixed starting threshold $\Theta_i = + 0.1\bar{\sigma}^r$ and different final threshold $\Theta_f$, ranging from $+0.5\,\bar{\sigma}^r$ to $+3.0\,\bar{\sigma}^r$.}
\vspace{-0.3cm} \label{tau_rally}
\end{figure*}
\indent We perform a similar analysis for stock price upturns or rallies. We find that the nonmonotonic behavior, with a maximum, of the MFHT vs volatility occurs both in real
market data and in simulations based on the proposed nonlinear Heston model (Eqs.~(\ref{NLH})-(\ref{CIR})). This means that we can extend our proposed measure of price
return stability not only to negative variations of price returns but also to positive variations. The results are shown in Fig.~\ref{tau_rally}. Again, we cross-validate the
results of Fig.~\ref{tau_rally}a by considering different thresholds with fixed interval size (Fig.~\ref{tau_rally}b), and fixed starting threshold with different final thresholds
(Fig.~\ref{tau_rally}c)~\cite{Note6}.\\
\indent This finding is similar to what is known to occur also in all physical systems with metastable states. Indeed, the behavior of the MFHT as a function of volatility shows the typical signature
of noise enhanced stability (NES) observed in a variety of physical, biological, chemical and ecological systems~\cite{Man96,Agu01,Dub04,DOd05,Hur06,Sun07,Yos08,Tra09,LiJ10,Val15}:
the stability of a metastable state can be enhanced by the noise and its average lifetime is a measure of this stability. This noise-enhanced metastability is a consequence of the interplay
between the thermal fluctuations and nonlinearity of the complex system investigated. This effect is observed by increasing the temperature as by analogy the stabilization of the price returns
occurs for increasing volatility. The empirical evidence of a NES effect in the behavior of the MFHT (Figs.~\ref{tau_crash} and~\ref{tau_rally}) suggests that price return dynamics could be
depicted by considering the value of the return as the position of a fictitious \emph{Brownian particle}, subjected to noise and moving in an effective potential with a metastable state.\\
\indent Models reproducing most of the stylised facts of financial markets and their dynamics by nonlinear stochastic differential equations, have been presented in literature~\cite{Bou03},
~\cite{Mal02,Hat06,Bou02}. Moreover, financial markets present different dynamical regimes with days of normal activity and days with large price variations, characterized by a different
behavior of volatility. In order to consider these different dynamical regimes and feedback effects on the price fluctuations, a Langevin approach to the market dynamics was already
proposed in Refs.~\cite{Bou03,Bou02,Bou98}, where a nonlinear stochastic dynamical equation with a metastable state was considered. The evolution inside the metastable state
represents the normal market behavior, while the escape from the metastable state represents the beginning of large price variations.\\
\indent Here, we employ a generalization of the Heston model~\cite{Hes93,Note7} proposed by~\cite{Bon07}, where the geometric Brownian motion is replaced by a random walk in the presence
of a cubic nonlinearity. This theoretical approach considers the financial market as an out-of-equilibrium system, whose dynamical evolution can be described by a nonlinear Heston model
defined by the following  It$\hat{o}$ stochastic differential equations~\cite{Gar04}
\begin{equation}
dx(t) = -\left(\frac{\partial U}{\partial x}+\frac{v(t)}{2}\right)dt + \sqrt{v(t)}\, dW_{1}(t), \label{NLH}
\end{equation}
\begin{equation}
dv(t) = a[b-v(t)]\, dt+c\,\sqrt{v(t)}\, dW_{2}(t), \label{CIR}
\end{equation}
with the volatility $v(t)$ given by the mean-reverting Cox, Ingersoll, and Ross (CIR) process~\cite{Fel51,Cox85,Hul11,Dra02,Note8}, and $U(x) = mx^{3} + nx^{2}$ is the effective
cubic potential with a metastable state (Fig.~\ref{tau-ret_pdf-pot}a). As explained in ~\cite{Bou98}, the parameters $m$ and $n$, identified in line with the consideration of liquid
markets, are influenced by the degree of risk aversion, the market depth, and the ``friction'' of prices to changes in demand and supply in the market. In Eq.~(\ref{NLH}), $x(t)=ln[p(t)/p(0)]$
is the return in the time window $[0,t]$, $p(t)$ is the price and $W_i$ are uncorrelated Wiener processes with $\langle dW_{i}(t)\rangle = 0$, and $\langle dW_{i}(t)\, dW_{j}(t')\rangle=dt\,\delta_{i,j}\delta(t-t')$.
We solve Eqs.~(\ref{NLH}) and (\ref{CIR}) numerically, obtaining a number of time series of returns equal to $1071$, with initial position $x_{0}=0.0$ and CIR stochastic process $v(t)$ defined
by $v_{start}=8.62\cdot10^{-5}$, $a=2.00$, $b=0.01$, and $c=0.83$~\cite{Note9}. Since we are focusing on the daily returns we have $x(t) \simeq r(t)$~\cite{Note10}. We fix again the two thresholds,
$\Theta_i = \mp 0.1\bar{\sigma}^r$ and $\Theta_f = \mp1.5\bar{\sigma}^r$ ($-$ for microcrash, $+$ for rally), where $\bar{\sigma}^r = 0.02383$ is the average standard deviation calculated over
the numerical time series. This yields, for the MFHT, the nonmonotonic behavior shown in Figs.~\ref{tau_crash}a and~\ref{tau_rally}a (red triangles), which exhibits a very close agreement with
the real data (blue circles).\\
\indent To quantitatively characterize the observed empirical results, we determine the probability distribution function (PDF) of the first hitting times of the daily returns, calculated by setting
$\Theta_i = - 0.1\bar{\sigma}^r$ and $\Theta_f = -1.5\bar{\sigma}^r$, and compare it with the corresponding theoretical PDF, obtaining a good qualitative agreement
(Fig.~\ref{tau-ret_pdf-pot}b)~\cite{Note11}.\\
\indent We then investigate the probability distribution of stock price returns, the PDF of volatility, the return correlation, and the absolute return correlation. We find that the agreement between
theoretical results and real data of all these statistical characteristics is quite good~\cite{Note12,Note13,Note14}. In particular, our model fulfills the property of the absence of return autocorrelation,
which ensures a no-arbitrage condition~\cite{Note14}. \\
\indent In summary, we have proposed using the MFHT as an indicator of price returns stability and looking at its relationship with returns volatility. In an empirical analysis
carried out on stocks traded at the NY Stock Exchange, the time series of daily returns show limited fluctuations, that is high stability, when volatility increases. In particular,
there is an intermediate range of volatility values where price returns show higher stability according to the proposed indicator. Moreover, the nonlinear Heston model
(Eqs.~(\ref{NLH})~-~(\ref{CIR})) appears to satisfy some of the well-established properties of financial markets and is able to reproduce the statistical properties of the hitting
times of daily returns in real stocks. The model is also able to describe the dynamics of price returns by considering an analogy between the metastability in the market and
that occurring in a variety of physical and complex systems~\cite{Pre11,Bon07},~\cite{Man96,Agu01,Dub04,DOd05,Hur06,Sun07,Yos08,Tra09,LiJ10,Val15}. Our findings
show that lower stability (smaller mean first hitting times) can be the result not only of large volatility, as it would be expected during periods of market ``turbulence''~\cite{Dat10},
but also of small volatility, which is usually considered an indicator of ``tranquil'' periods. This result could bear important implications both for practitioners and policy-makers
responsible for market stability. Further, the proposed measure can be considered as an additional useful indicator to monitor market stability.\\
\indent It is worth mentioning that the \emph{clustering} phenomenon of volatility is important for understanding the instabilities in price returns and the nonmonotonic behavior
of the MFHT vs volatility~\cite{Note15}.\\
\indent Finally we note that the applications of our definition of stability based on the concept of first hitting time can help to quantitatively characterize the resilience of different
complex systems, both in physics and biology (such as in neuronal activity and population dynamics), to variations of a given feature.

\vspace{-0.21cm}
\section{Acknowledgments}

\vspace{-0.21cm}
\indent We gratefully acknowledge Rosario N. Mantegna for helpful discussions and critical reading of the manuscript, and the Observatory of Complex Systems that provided
us the real market data used for our investigation.


\begin{thebibliography}{30}

\bibitem{Min92}
P. H. Minsky, \emph{The Financial Instability Hypothesis}, Levy Economics Institute Working Paper No.74 (1992).

\bibitem{Yel14}
J. Yellen, Transcript of Chair Yellen's Press Conference, June 18, (2014). https: //www.federalreserve.gov/media\\center/files/FOM C presconf20140618.pdf

\bibitem{Man95}
R. N. Mantegna, H. E. Stanley, Nature \textbf{376}, 46 (1995).

\bibitem{Man00}
R. N. Mantegna, H. E. Stanley \emph{An Introduction to Econophysics: Correlations and Complexity in Finance} (Cambridge Univ. Press, 2000).

\bibitem{Bou03}
J. P. Bouchaud, M. Potters \emph{Theory of Financial Risks and Derivative Pricing} (Cambridge University Press, 2003).

\bibitem{Ple03}
V. Plerou, P. Gopikrishnan, H. E. Stanley, Nature \textbf{421}, 130 (2003).

\bibitem{Lil03}
F. Lillo, J. D. Farmer, R. N. Mantegna, Nature \textbf{421}, 129 (2003).

\bibitem{Bou08}
J. B. Bouchaud, Nature \textbf{455}, 1181 (2008).

\bibitem{Yak09}
V. M. Yakovenko, Rev. Mod. Phys. \textbf{81}, 1703 (2009).

\bibitem{Wan08}
F. Wang, K. Yamasaki, S. Havlin, H. E. Stanley, Phys. Rev. E \textbf{77}, 016109 (2008).

\bibitem{Adr14}
T. Adrian, D. Covitz, \& N. Liang, Financial Stability Monitoring, \emph{Federal Reserve Bank of New York Staff Reports no. 601}, pp. 1--50 (2014).

\bibitem{Eng04}
R. Engle, Am. Econ. Rev. \textbf{94}, 405 (2004).

\bibitem{Bia09}
S. Bianco, F. Corsi, R. Ren\`{o}, Proc. Natl. Acad. Sci. USA \textbf{106}, 1439 (2009).

\bibitem{Yam05}
K. Yamasaki, L. Muchnik, S. Havlin, A. Bunde, H. E. Stanley, Proc. Natl. Acad. Sci. USA \textbf{102}, 9424 (2005).

\bibitem{Pod09}
B. Podonik, D. Horvatic, A. M. Peterson, H. E. Stanley, Proc. Natl. Acad. Sci. USA \textbf{106}, 22079 (2009).

\bibitem{Pre11}
T. Preis, J. J. Schneider, H. E. Stanley HE (2011), Proc. Natl. Acad. Sci. USA \textbf{108}, 7674 (2011).

\bibitem{Moh11}
B. Mohr, H. Wagner, A structural approach to financial stability, \emph{Discussion Paper No. 467, May 2011, FernUniversit\"{a}t in Hagen}, pp. 1--42 (2011).

\bibitem{Dat10}
P. Dattels, R. McCaughrin, K. Miyajima, J. Puig, Can you map global financial stability? \emph{IMF Working Paper, WP/10/145, International Monetary Fund}, pp. 1--43 (2010).

\bibitem{Gad09}
B. Gadanecz, K. Jayaram, Measures of financial stability - a review, \emph{IFC Bulletin No. 31, Bank for international settlements}, pp. 1--16 (2009).

\bibitem{And09}
T. G. Andersen, D. Dobrev, E. Schaumburg, Duration-based volatility estimation, \emph{Global COE Hi-Stat Discussion Paper Series 034, Institute of Economic Research
Hitotsubashi University} pp. 1-66 (2009).

\bibitem{Bon07}
G. Bonanno, D. Valenti, B. Spagnolo, Phys. Rev. E \textbf{75}, 016106 (2007). Here, the dynamical regimes corresponding to different parameter values were studied and
the parameter region was found to observe a nonmonotonic behavior of the MFHT.

\bibitem{Mas08}
J. Masoliver, J. Perell\'{o}, Phys. Rev. E \textbf{80} 016108 (2009); Phys. Rev. E \textbf{78}, 056104 (2008).

\bibitem{Note1}
See Sec. I, Fig.~S1 and Refs~S1-S6 of the Supplemental Material.

\bibitem{Note2}
The mean first hitting time (MFHT) or the mean first passage time (MFPT), and the problem of the random walk with reflecting and absorbing barriers was earlier introduced
in scientific literature, see the following referenes~\cite{Kra40,Cha43,Fel51}. In finance, the concept of first-passage time (FPT) appears in several domains: valuation of barrier
options, credit risk modeling and optimal exercise time of American options. Even more, the study of the MFPT for two well-known mean-reverting processes, that is the square
root process of Feller and the GARCH diffusion process, was recently given in Refs.~\cite{Zha10,Mas12}.

\bibitem{Kra40}
H. A. Kramers, Physica \textbf{7}, 284 (1940).

\bibitem{Cha43}
S. Chandrasekhar, Rev. Mod. Phys. \textbf{15}, 1 (1943)

\bibitem{Fel51}
W. Feller,  Ann. Math. \textbf{54}, 173 (1951).

\bibitem{Zha10}
Bo. Zhao, \emph{Mean first-passage times of the Feller and the GARCH diffusion processes},
 https://urlsand.esvalabs.com/?u=http
City University London - Sir John Cass Business School. Cass Business School, London (2010).

\bibitem{Mas12}
J. Masoliver, J. Perell\'{o}, Phys. Rev. E \textbf{86} 041116 (2012).

\bibitem{Note3}
These stocks, readily available in the common financial data sources, have been continuously recorded for twelve years from $1987$ to $1998$
($3030$ trading days) and therefore represent the overall market performance over a reasonably long time span. Moreover, this dataset has been
used in previous investigations (see Refs.~[S18, S19] of the Supplemental Material and next Refs~\cite{Mic02,Bon03,Bon04}).

\bibitem{Mic02}
S. Miccich\`{e}, G. Bonanno, F. Lillo, R. N. Mantegna, Physica \textbf{314} 756 (2002).

\bibitem{Bon03}
G. Bonanno, G. Caldarelli, F. Lillo, R. N. Mantegna, Phys. Rev. E \textbf{68} 046130 (2003).

\bibitem{Bon04}
G. Bonanno, G. Caldarelli, F. Lillo, S. Miccich\`{e}, N. Vandewalle, R. N. Mantegna, Eur. Phys. J. B \textbf{38} 363 (2004).

\bibitem{Note4}
To assess the robustness of the results we consider a wide range of realistic numerical parameters for $\theta_i$ and $\theta_f$
in the intervals $[+0.9, -1.6]$ and $[-0.5, -3.0]$, respectively in order to identify an episode of instability~\cite{Eic95,Faz07}.

\bibitem{Eic95}
B. Eichengreen \emph{et al.}, Econ. Policy \textbf{21}, 249 (1995).

\bibitem{Faz07}
G. Fazio, J. Int. Money Financ. \textbf{26}, 1261 (2007).

\bibitem{Note5}
Considering the MFHT as a measure of market stability, it is possible to argue that volatility plays a stabilizing effect
when its values are within the range $[0.004, 0.01]$.

\bibitem{Note6}
The coalescence of the curves of MFHT vs volatility for fixed difference between thresholds (see Fig.~\ref{tau_rally}b) could be ascribed to the
asymmetry of the returns distribution, characterized by a negative skewness (see Fig.~S2 of the Supplemental Material).

\bibitem{Man96}
R. N. Mantegna, B. Spagnolo, Phys. Rev. Lett. \textbf{76}, 563 (1996).

\bibitem{Agu01}
N. V. Augdov, B. Spagnolo, Phys. Rev. E \textbf{64}, 035102(R) (2001).

\bibitem{Dub04}
A. A. Dubkov, N. V. Agudov, B. Spagnolo, Phys. Rev. E \textbf{69}, 061103 (2004).

\bibitem{DOd05}
P. D'Odorico, F. Laio, L. Ridolfi, Proc. Natl. Acad. Sci. USA \textbf{102}, 10819 (2005).

\bibitem{Hur06}
P. I. Hurtado, J. Marro, P. L. Garrido, Phys. Rev. E \textbf{74}, 050101(R) (2006).

\bibitem{Sun07}
G. Sun, Dong N., Mao G., Chen J., Xu W., Ji Z., Kang L., Wu P., Yu Y., Xing D., Phys. Rev. E \textbf{75}, 021107 (2007).

\bibitem{Yos08}
M. Yoshimoto, H. Shirahama, S. Kurosawa, J. of Chemical Phys. \textbf{129}, 014508 (2008).

\bibitem{Tra09}
M. Trapanese, J. Appl. Phys. \textbf{105}, 07D313 (2009).

\bibitem{LiJ10}
J. H. Li, J. Luczka, Phys. Rev. E \textbf{82}, 041104 (2010).

\bibitem{Val15}
D. Valenti, L. Magazz\`u, P. Caldara, \& B. Spagnolo, Phys. Rev. B \textbf{91}, 235412 (2015).

\bibitem{Mal02}
O. Malcai, O. Biham, P. Richmond, S. Solomon, Phys. Rev. E \textbf{66}, 031102 (2002).

\bibitem{Hat06}
J. P. L. Hatchett, R. K\"{u}hn R, J. Phys. A \textbf{39}, 2231 (2006).

\bibitem{Bou02}
J. P. Bouchaud, Physica A \textbf{313}, 238 (2002).

\bibitem{Bou98}
J. P. Bouchaud, R. Cont, Eur. Phys. J. B \textbf{6}, 543 (1998).

\bibitem{Hes93}
S. L. Heston, Rev. Financ. Stud. \textbf{6}, 327 (1993).

\bibitem{Note7}
See Sec. II of the Supplemental Material.

\bibitem{Gar04}
C. W. Gardiner {\em Handbook of Stochastic Methods}, (Springer, Berlin, 2004).

\bibitem{Cox85}
J. C. Cox, J. E. Ingersoll, and S. A. Ross, Econometrica \textbf{53}, 385 (1985).

\bibitem{Hul11}
J. C. Hull, \emph{Options, Futures, and Other Derivatives} (Prentice Hall, London, 2011).

\bibitem{Dra02}
A. A. Dr\v{a}gulescu, V. M. Yakovenko, Quant. Finance \textbf{2}, 443 (2002).

\bibitem{Note8}
See Sec. II and Refs.~S14-S17 of the Supplemental Material.

\bibitem{Note9}
These values were obtained by best fitting between theoretical and empirical results for all the statistical features investigated, by performing both $\chi^2$
and Kolmogorov-Smirnov (K-S) goodness-of-fit tests (Figs.~\ref{tau_crash}-\ref{tau-ret_pdf-pot}b, and Figs.~S2 and~S3 of the Supplemental Material).

\bibitem{Note10}
For the daily returns we have that: \\ $x_d(t) = ln (p(t)/p(0)) - ln (p(t-1)/p(0)) = ln(p(t)) - ln(p(t-1))
\simeq  (p(t) - p(t-1))/(p(t-1)) = r(t)$, where we use $ln \thinspace x \simeq x - 1$.

\bibitem{Note11}
Performing both $\chi^2$ and Kolmogorov-Smirnov (K-S) goodness-of-fit tests, we get: $\chi^2\thinspace=\thinspace0.01668$,
$\tilde{\chi}^2\thinspace=\thinspace0.00018$ (reduced $\chi^2$) and $D=0.149$, $P=0.198$. $D$ and $P$ are respectively the maximum difference between
the cumulative distributions and the corresponding probability for the K-S test. The results indicate that the two distributions are not significantly different.

\bibitem{Note12}
See Figs.~S2 and~S3 of the Supplemental Material. In Fig.~S2 we show the experimental and theoretical PDFs of the returns. The agreement between theoretical results and
real data is quite good, except at high values of the returns. This can be ascribed to the failure of the proposed nonlinear model for returns higher or comparable to the height
of the metastable state barrier. For these values of returns other mechanisms, which we have not taken into account in the model, come into play~\cite{Bon07,Bou98}.

\bibitem{Note13}
The quantitative statistical characterization of the shape of the PDF of returns (see Sec. III of the Supplemental Material) shows that the model reproduces the asymmetry and
leptokurtic distribution observed for the real market data~\cite{Man00,Bou03}, and the PDF of the volatility, both for real market data and theoretical results, shows a log-normal
behavior (see Fig.~S3a of the Supplemental Material).

\bibitem{Note14}
In Fig. S3b of the Supplemental Material, the autocorrelation of the asset returns and absolute return correlation, obtained both from theoretical calculations and real data,
are shown. The autocorrelations from the model (red triangles in Fig.~S3b) are insignificant except for very short times where microstructure effects possibly come into play.
This result is again in close agreement with the autocorrelations calculated for the real data (blue circles in Fig.~S3b). A similar behavior, but with a slow decay
to zero, is displayed by the correlation function of the absolute returns (see the inset of Fig.~S3b).

\bibitem{Note15}
Indeed, the contemporaneous presence, in the time series of returns, of pairs of ``\emph{clustering}'' and/or ``\emph{spikes}'' spaced  from a nearly laminar or ``tranquil'' regime
gives rise to a nonmonotonic increase of the MFHT, with a presence of a maximum, in the intermediate region of volatility values (see the ending of Sec. III of the Supplemental Material).



\end{thebibliography}
\end{document}